\documentclass[12pt]{article}
\begin{document}

\def\be{\begin{equation}}
\def\ee{\end{equation}}
\def\bea{\begin{eqnarray}}
\def\eea{\end{eqnarray}}
\def\nzam{non zero angular momentum component}
\def\myp{Myers-Perry}

\begin{titlepage}
\begin{center}
\hfill hep-th/0205087\\  
\hfill IITK/PHY/2002/36\\  

\vskip .2in

{\Large \bf Rotating Black Holes in Higher Dimensional 
Brane Worlds}
\vskip .5in
{\bf Gautam
Sengupta }\\
\vskip .1in
{\em Department of Physics,\\
Indian Institute of Technology\\
Kanpur 208 016\\
INDIA}
\vskip .5cm

\end{center}
\begin{center} {\bf ABSTRACT}
\end{center}
\begin{quotation}\noindent
\baselineskip 10pt

A black string generaliztion of the Myers-Perry N dimensional
rotating black hole
is considered in an (N+1) dimensional
Randall-Sundrum brane world. The black
string intercepts the (N-1) brane in a N dimensional rotating black
hole. We examine the diverse cases arising for various
non-zero rotation components and obtain the geodesic equations for these
space-time. The asymptotics of the resulting brane world geometries and
their implications are discussed.

\end{quotation}
\vskip .2in
May 2002\\
\end{titlepage}

\section {Introduction.}

String theory and the unification of gravity with the
other fundamental
forces have led to the consideration of higher dimensional space-time 
at ultrashort length scales. It has been generaly 
assumed that the extra spatial dimensions are small and compactified at
the Planck
scale. Recent investigations have however led to the possibility 
of consistent models with large compact dimensions also.
This however
required the assumption that the electroweak scale was the
fundamental scale and equal to the higher 
dimensional Planck scale. The large four
dimensional Planck scale was then a derived scale enhanced by the (large)
volume
factor. Furthermore such a scenario required that the
gauge
sector of the fundamental interactions are restricted
on a three brane (or a
smooth four dimensional hypersurface) \cite{add}. This
{\it brane world} model removed the
hierarchy between
the electroweak and the Planck scale. However it was
reintroduced 
in a
hierarchy of scales between the large and the small dimensions. 
Nonetheless, these models
inspired the exciting phenomenological possibilty of low scale
quantum gravity which may be accessible in the next generation
accelarators. Subsequently Randall and Sundrum (RS) \cite{rs12} considered
a
five dimensional model with a metric involving {\it warped
compactification} and the
standard model interactions being restricted on a three brane,  as a
resolution to the
hierarchy problem. This model also
required a regulator three
brane located at a certain distance from the first one
in the extra fifth spatial direction. In this case the hierarchy of scales
are absent as the compactified directions remained small in length.
In a subsequent variant the regulator brane could be shifted an infinite
distance away leading to a model with a single three brane and a large
( non-compact) extra fifth direction not sensitive to the gauge
interactions. The gravitational interactions however were free to
propagate in all the dimensions.
For such a model to be an acceptable
solution of the five dimensional Einstein equation the corresponding
space-time was required to be a slice of a Anti deSitter space in five
dimensions with reflection symmetry at the location of the three brane. 
In this framework the zero mode
of the Kaluza Klein graviton was localized on the three brane in a
linearized approximation
and acted as the source for the usual weak four dimensional
gravity. Last couple of years have witnessed intense investigations
in this area leading to a large number of insights into the
exciting consequences
of these interesting models. It was found that the conclusions of
the linearized analysis carried over identicaly to any solution
with a Ricci flat metric.
A full non linear treatment in the framework of supergravity
\cite{gibs} have confirmed the conclusions resulting from the linearized
approximation.  In particular, these constructions have led to the
possibility of detecting low scale quantum gravity effects on phenomenlogy
at the weak scale in the next generation particle acclerators \cite{lyk}.  

For these models to be consistent, General Relativity and its conclusions
in four dimensions should be reproduced. In particular it should be
possible to obtain
cosmological and
black hole solutions in four dimensions starting from such higher
dimensional scenarios. Study of black hole solutions have been
an exciting aspect of investigations in this area.
Such a four dimensional black hole is expected to be extended in the extra
spatial AdS
direction and
would thus be a higher dimensional object in the brane world. 
In an interesting article Chamblin, Hawking
and Reall ( CHR)\cite {chr} considered the description of a Schwarzschild
black hole
on the three brane in a RS brane world. The bulk solution in this case
was a
five dimensional black string. This solution
showed the usual Schwarzschild singularity on the three brane in a
linearized framework and was
extended along the transverse fifth direction. Thus it described a
usual four dimensional Schwarzschild black hole on the three brane.
It was observed however
that the solution was singular at the AdS horizon with the
singularity
there being a p-p curvature singularity \cite {pps,gibs}. It was argued
that the solution
was
unstable far away from the brane due to the Gregory-Laflamme instability
\cite {greg1}
and was conjectured to pinch off before reaching the horizon to lead to
a cigar like solution. It was later shown that for these AdS solutions
it was more likely that there would be an accumulation of mini black holes
towards the AdS horizon which did not seem to indicate a cigar geometry.
Attempts to consider the off-brane metric in the linearized
framework have also appeared in \cite {gian}. Further generaliztions of
these ideas have appeared in \cite {exrefs}.
It must however be stated here that the conclusions of Gregory and
Laflamme
and hence their consequences on the brane world solutions
have been questioned in \cite{maeda}. There it has been shown under
very mild assumptions that
pinching off of the horizon for a Schwarzschild black string was
untenable from stability considerations. It was suggested there that 
a 
more likely scenario would be that the system will evolve to an
translationaly non-invariant intermediate stable
solution. In a subsequent article such a solution 
has been alluded to \cite{maeda} in the form of the time symmetric
initial data. However the argument has not yet been
generalized to
include the axisymmetric solutions also.

Generalizations of this construction for
charged
black holes and numerical studies for the off-brane bulk metric has also
been performed \cite{kim},\cite {cbh},\cite {num}. Exact studies of
black
holes in lower dimensional brane
worlds have been undertaken in the context of the AdS C metric for 2
branes embedded in a ( 3 + 1) dimensional RS brane world \cite{emp}.
Rotating
generalizations of these constructions both for the Kerr like and the
BTZ variants have also been discussed. It was observed that the
corresponding
black hole solutions were non-singular at the horizon which possibly
indicates that the singularity at the AdS horizon is an artifact of the
linearized analysis. However such exact C-metric solutions are
unavailable
in the realistic four or higher dimensions.
In an earlier article we have generalized the construction of CHR to
rotating black holes on the brane
in a five dimensional RS brane world \cite {gs}. This was obtained by
considering
a five dimensional rotating black string solution
which intercepted the three brane in a four dimensional Kerr black hole.
It was found that the Kerr solution too was singular at the AdS horizon
apart from the usual ring singularity at $r=0$ on the brane.
Study of the geodesic equations then showed that it was plausible that the
singularity at the AdS horizon in the rotating case was also a p-p
curvature singularity although an explicit construction of the parallely
propagated orthonormal frame was computationaly intractable.

It must be understod that for consistency the RS model ( or some variant )
must be
embedded as a low energy description in an appropriate string theory.
Thus it requires a generalization of such models to higher dimensions.  
Generalization of the Randall-Sundrum construction to arbitrary
(N+1) dimensions
with an appropriate co-dimension 1 brane is straightforward and may be
easily extended
to include the full non-linear equations for a Ricci flat metric
\cite{emp}.  Furthermore several variants of the RS scenario involving
cosmic strings and other global defects of various co-dimensions
and their
consistency has been investigated in higher dimensions \cite {hd}. Just
as in five dimensions, the conclusions of General Relativity on the
appropriate co-dimension brane needs to be reproduced for consistency.
In particular it should be possible to obtain known higher dimensional
black hole solutions starting from such a ( RS like) scenario. The absence
of exact C-metrics in dimensions $D\geq 3$ requires such studies to be
based on the linearized approximation and to that end the CHR model
is a reasonable approach illustrating the physical issues involved 
although the obvious problem of boundary singularities persist.

In this article we consider the description of higher
dimensional black holes on a codimension 1 brane embedded in a (N+1)
dimensional RS brane world with a single AdS direction. In particular
we consider the emergence of a generalization of the Kerr solution to 
N dimensions due to Myers and Perry \cite{myers} on the (N-1)
dimensional brane. 
The corresponding case for the non-rotating higher dimensional
Schwarzschild black hole due to Tangherlini et. al \cite {tangh}
from a higher dimensional brane world is a trivial
extension of the CHR construction and we do not explicitly consider it
here.
The
Myers-Perry 
solution however shows rather interesting behaviour in the fact
that in N
dimensions the symmetry group of the space-time is $SO(N-1, 1)$ which
admits multiple rotations in the various coordinate planes
consistent with the number of the Casimir operators. Furthermore these
solutions show a variety of horizon and singularity structures depending
on whether the number of spatial dimensions are even or odd. So it is an
interesting exercise to describe these structures from a brane world
perspective apart from the motivations outlined earlier.
To this end
we have considered a rotating black string in the (N+1) dimensional brane
world.
This black string intercepts the (N-1) brane in a N dimensional rotating
black hole solution. We discuss the various cases of horizon and
singularity structures with non zero angular
momentum parameters and even or odd number of spatial dimensions.
We also obtain
the geodesic equations for the most general case and discuss their
consequences on the
asymptotics of
these space-time. The article is organized as follows, in the next section
we briefly review the Myers and Perry solution for both single and
multiple non zero
angular
momentum components. In section three we consider a black string
generalization of the Myers-Perry solution with a single non-zero angular
momentum
component in a (N+1)dimensional RS brane world. In the section following
that we
treat the most general case of a rotating black string with multiple
angular momentum components in a (N+1) dimensional RS brane world. In the
last section we present a summary and our conclusions.

\section {Myers-Perry Solution in N Dimensions }

In this section we briefly review the Myers-Perry solution \cite
{myers} for a rotating
black
hole in N dimensions. In (3 + 1) dimensions the rotating
black hole described by a Kerr metric is characterized by the two
parameters mass and the
angular momentum invariant. In N ( $N > 4$ ) dimensions the
Poincare group contains the $ SO(N-1, 1)$ Lorentz group..
For a massive representation of this group the Casimir invariants are
the mass and the $[\frac{N-1}{2}]$ Casimirs of the little group $SO(N-1)$.
Hence a rotating black hole in N dimensions is characterized by 
$[\frac{N-1}{2}] +1$ parameters being the mass and corresponding angular
momentum invariants. 

We first consider the case of a single angular momentum of the
N dimensional black hole.
For a single \nzam the metric in a Boyer-Lindquist coordinate
system is given as
\bea
ds^2 &=& \big[ -(1-{\mu\over {r^{N-5}\rho^2}})dt^2 + ( r^2 + a^2
+{{\mu a^2}\over{r^{N-5}\rho^2}} \sin^2 \theta )sin^2 \theta d\phi^2
\nonumber \\
&+& {\rho^2\over\Delta} dr^2 + \rho^2d\theta^2 + r^2 \cos^2\theta
d\Omega^{N-4}
 - {{2 \mu a}\over{ r^{N-5}\rho^2}}d\phi dt \big ]
\eea
where $\rho^2=r^2 + a^2 \cos^2 \theta$ and $d\Omega^{N-4}$ is the line
element on a unit ( N-4) sphere, where for N=4 the last term is absent.
The 
function $\Delta={\mu\over {r^{N-5}\rho^2}}$ and M is the
ADM mass of the black hole which is given as 
$M=\frac{(N-2)A_{N-2}}{16 \pi}\mu $
and
$A_{N-2}=\frac{2\pi {\frac{N-1}{2}}}{\Gamma ({\frac{N-1}{2}})}$
is the area of the unit ( N-2) sphere. This solution posesses only
one non-zero angular momentum component in the $(x,y)$ plane in a
quasi spheroidal $(x,y,z,t)$ coordinates in the flat space limit 
$(\mu \rightarrow 0)$, given as \cite {myers}
\be
J^{yx}={A_{N-2}\over {8 \pi}}\mu a = {2\over {N-2}}M a
\ee
The horizon occurs at $g^{rr}=0$ and a solution always exists for $N > 5$
irrespective of the angular momentum unlike that in four dimensions.

However this result is not the most general, as the non-zero components
of the angular momentum will have two spatial components indicating the
specific coordinate planes for the rotation. This is related to the
number of Casimir invariants of $SO(N-1)$ or the dimension of its Cartan
subalgebra which generates commuting rotations in the corresponding
coordinate planes. In the metric presented in eqn. (1) there is only one
such non-zero
angular momentum component. The general solution will have 
$[\frac{N-1}{2}]$ 
such non-zero components along the various coordinate planes. The general
metric for both odd and even
(N-1) spatial coordinates is given as \cite {frolov}
\bea
ds^2 &=& -dt^2 +\sum_k (r^2 + a_k^2)(d\mu_k^2 + \mu_k^2d\phi_k^2)
+ {{\mu r^{2-\epsilon}}\over \Pi F}\big [ dt - \sum_k a_k\mu_k^2
d\phi_k\big ]^2 \nonumber \\
&+& {{\Pi F}\over {\Pi -\mu r^{2-\epsilon}}} dr^2
+\epsilon r^2 d\alpha^2,
\eea
where $\epsilon =0,1$ for odd/even N and the sums extend over
$k={{N-2}\over 2}$ for N even and $k={{N-1}\over 2}$ for N odd. The
parameters
in the metric are given as
\be
\mu={{16\pi}\over {(N-2)A_{N-2}}} M;\ \ a_k= {{N-2}\over 2}{J_k\over M};\
\ 
\sum \mu_k^2 + \epsilon \alpha^2=1.
\ee
Here $\phi_k$ are angles in each of the planes $(x^k, y^k)$ and $\mu_k$
are
direction cosines $(0\leq \mu_k\leq 1)$ with respect to these planes. The
functions $\Pi$ and $F$
are given as
\be
F= 1 -\sum_{k} {{a_k^2\mu_k^2}\over {r^2 + a_k^2}},\ \ \ \Pi=\prod_k (r^2
+
a_k^2 ).
\ee
The metric admits the following Killing isometries $\partial_t$ and
$\partial_{\phi_k}$ for translations in $t$ and rotations in
$\phi^k$ justifying its stationary nature in N dimensions.

The Myers-Perry solution presented above shows interesting horizon and
singularity structures depending on the number of non-zero angular
momentum components and whether N is odd or even. It is useful to check
the surfaces of constant $r$ for these solutions in order to clarify
their global characteristics. For odd (N-1) spatial
components , from a Kerr-Schild form for the metric \cite {myers}
we have the equation
\be
\frac{x^{i^2} + y^{i^2}}{r^2 + a_i^2} + \frac{w^2}{r^2}=1.
\ee
Here $(x^i,y^i)$ are the paired spatial coordinates defining the
coordinate planes for the rotation components and $w$ is the
unpaired one with $\phi_i$ being
the
angles in the respective coordinate planes defined by the pairings
$(x^i,y^i)$. Eqn. (6) defines a family of (N-2)
ellipsoids in $R^{N-1}$ parametrized by $r$. The intersection of these
with $x^i=y^i=0$ gives the familiar two dimensional surfaces of a Kerr
metric. These are essentialy ellipsoids of revolution about the $w$ axis
or $S^2$ squashed in the $w$ direction. They intersect the $w$ axis at
$w=r$
and the $(x^1, y^1)$ plane at the circle $x^{1^2} + y^{1^2}=r^2 + a_1^2$
so the $r=0$
surface degenerates to the disk $x^{1^2} + y^{1^2}\leq a_1^2$ in the
$(x^1,
y^1)$ plane. The entire surface may be described as a squashed ( N-2)
sphere with rotational symmetry in each  $(x^i,y^i)$ plane and the $r=0$
surface degenarates to an $(N-2p -2)$ ball where $p$ is the number of
vanishing rotation parameters $a_i$. An M ball is defined as a region of
$R^M$ with an (M-1) sphere as boundary.
For even (N-1) number of spatial
coordinates with the unpaired coordinate $w$ absent,
we have from similar considerations in the Kerr-Schild form \cite {myers}
\be
{{x^{i^2} + y^{i^2}}\over {r^2 + a_i^2}}=1
\ee
This describes an ellipsoid again with $r=0$ surface being an $(N-2p-1)$
ball
if $p\neq0$. If $p=0$ then no surfaces of constant r intersects the
origin.

The horizon for these metrics in the Boyer-Lindquist coordinate system
requires analytic solutions for the equation,
\be
\Pi -\mu r^{2-\epsilon} =0
\ee
where $\epsilon$ takes values $(0, 1)$ as mentioned earlier.
For odd (N-1) spatial coordinates with no vanishing rotation parameters
$a_i$ it was shown that if an horizon exists then it has the topology of
$S^{N-2}\times R$ with $r$ positive and three possible cases
\bea
\Pi -\mu r &>& \ 0 \ \  \mbox{ no horizon} \nonumber\\
           &<& \ 0  \ \ \mbox {two horizons}\nonumber\\
           &=& \ 0  \ \ \mbox {one degenerate horizon}
\eea
as in the familiar Kerr case in four dimensions. It was found that
beside N=4 the equation only has a solution for N=6 though
there was a possibility of solutions for larger N. The
vanishing
of at least one rotation parameter was found to be a sufficient condition
for the
existence of a horizon. For even (N-1) spatial coordinates one requires
the condition
\be 
\Pi -\mu r^{2} =0
\ee
Once again an analytic solution for arbitrary N is difficult. For N=5
however this is just a quadratic equation in $r^2$ with the solution
\be
2r^2_{\pm}=\mu -a_1^2 -a_2^2 \pm \sqrt {(\mu -a_1^2 -a_2^2) - 4a_1^2a_2^2}
\ee
so the existence of a horizon requires 
\be
\mu\geq a_1^2 + a_2^2 + 2\mid a_1 a_2 \mid
\ee
For arbitrary N it was found that analytic solutions were possible for N=7
,9, 11. In this cases also it was observed that there were three cases as
in eqn. (9) for (N-1) odd but a single vanishing rotation parameter was
insufficient
to guarantee the existence of a horizon in this case. It needed at least
two such non-zero parameters. Like the four dimensional Kerr solution
the Myers-Perry metric too posesses an ergosphere or an infinite redshift
surface
where $g_{tt}=0$. The outer boundary of this is the  
stationary limit surface which satisfies the equation
\be 
\Pi F -\mu r^{2-\epsilon} =0
\ee
and has the topology of $S^{N-2}\times R$. 

The singularities of these metrics for both odd and even ( N-1) spatial
coordinates are best examined in the Kerr-Schild coordinates. The analysis
in \cite {myers} shows that the metric is smooth everywhere except where
$h=\frac {\mu r^{2-\epsilon}}{\Pi F}$ diverges. For odd ( N-1 ) spatial
coordinates
we may rewrite 
\be
F=\sum_{i}\frac{r^2\mu_i^2}{r^2 + a_i^2} + \alpha^2
\ee
where $\alpha={w\over r}$.
In this case if any of the rotation parameters $a_i=0$ then $\Pi$ involves
factors of $r^2$
causing $h$ to diverge as $r=0$. If all the $a_i\neq 0$ then the only
singularities arise from $F=0$ which requires $\alpha=\frac{w}{r}=0$.
Then $F$ contains factors of $r^2$ which causes $h$ to diverge as $r=0$.
The singularity occurs at $ w=0, \ \ \ \frac{x^{i^2} + y^{i^2}}{a_i^2}=1 $
which is the edge of the (N-2) ball at $r=0$ whereas $h$ remains finite in 
the interior. This is analogous to the ring singularity of the four 
dimensional Kerr metric. For even ( N-1) spatial coordinates $h=\frac {\mu
r^2}{\Pi F}$ and $F$ may be expressed as $F=\sum_i\frac{r^2\mu_i^2}{r^2 +
a_i^2}$. For all $a_i\neq 0$, $\Pi$ is finite everywhere and $F$ contains
a factor $r^2$ which cancels that in numerator of $h$. So $h$ is finite
except at the origin which is excluded. If one spin parameter $a_1=0$
then $\Pi$ has a single $r^2$ factor which cancels that in numerator of
$h$. Then for $F$ to vanish it requires $\mu_1=0$ which means $x^1=y^1=0$.
Then $F$ contains factors of $r^2$ which causes $h$ to diverge as $r=0$
but only at the edge of the ( N-3) ball $x^1=y^1=0, \sum_{i > 1}\frac
{x^{i^2}+y^{i^2}}{a_i^2}=1$. this is once again the analog in higher
dimensions of the usual ring singularity. 
To establish if these were true singularities of the space-time geometry
the components
of the curvature tensor were evaluated \cite {myers} in an orthonormal
frame 
indicating the tidal forces. A specific component 
$ R_{uvuv} \sim r^{-2p-\epsilon }$ and diverged for most cases.
However certain exceptional cases were listed in \cite {myers} where $r=0$
does 
not appear to be entirely singular and extension to negative $r$ seemed
necessary. We do not consider those specificaly over here.
Given the Myers-Perry solution
it is straightforward to obtain the corresponding black string/brane 
generalization
in a higher dimensional space-time by adding extra flat directions
to the metric. In the black string case the horizon will now have the
topology of $S^{N-2}
\times R^1 \times R^1$ and an extended singularity along the extra
flat direcion.

\section {Single Spin Myers-Perry Black String in a Brane World}

In this section after a brief review of an (N + 1) dimensional RS brane
world, we consider a black string version of the Myers-Perry solution
for a single non-zero rotation parameter in the higher dimensional bulk.
It can then be shown that this leads to a N dimensional rotating black
hole metric of \myp\  on the (N -1) brane world volume. The RS brane world
in
(N+1) dimensions
may be described by the metric \cite {emp} 
\be
ds^2=g_{mn}dx^mdx^n={l^2\over z^2}\big [ g_{\mu\nu}dx^{\mu}dx^{\nu} + dz^2
\big ].
\ee
Here  $\mu,\nu = 1..N $ 
and $l$ is the AdS length scale and $z=0, \infty$ the conformal infinity
and the AdS horizon respectively, $z$ being the direction transverse to
the N-1 brane. 
The actual RS geometry is obtained by removing the small $z$ region at
$z=z_0$ and glueing a copy of the large $z$ geometry. The
resulting topology is
essentialy $R^N \times {{S^1}\over {Z_2}}$.
The discontinuity 
of the extrinsic curvature at
the $z=z_0$ surface corresponds to a thin distributional source of stress-
energy. From the Israel junctions conditions this may be interpreted as
a relativistic (N-1) brane with a corresponding tension \cite{gid, emp}.
Another variant of this model is to slice the AdS space-time
both at $z=0$ and $z=l$ and insert two (N-1) branes with $Z_2$ reflection
symmetry at both the surfaces. The Israel conditions now require a
negative tension for the brane at $z=l$. The first version may be obtained
from the second by allowing the negative tension brane to approach the
AdS horizon at $z=\infty$ however a dynamical realization of this is not
yet clear. We will focus our considerations on the first
variant of the RS geometry in this case but state that our construction
may be generalized to the second variant also.
The Einstein equations in N +1 dimensions with a negative cosmological
constant continue to be satisfied for any metric $g_{\mu\nu}$ which is
Ricci flat. The curvature of the modified metric now satisfies
\be
R_{pqrs}R^{pqrs}= {{2N(N+1)}\over l^4} +{z^4\over
l^4}R_{\mu\nu\lambda\kappa}R^{\mu\nu\lambda\kappa}
\ee where $(p,q)$ runs over (N +1) dimensions.
The perturbations of the N dimensional metric are now normalizable modes
peaked at the location of the (N-1) brane.

CHR \cite{chr} considered the description of a Schwarschild black hole
on a three brane in the corresponding five dimensional RS brane world.
The metric on the brane
in this case must be a Schwarzschild metric which is Ricci flat. However
the obvious choice of a five-dimensional AdS-Schwarzschild metric failed 
to
satisfy the Israel junction conditions compatible with the $Z_2$
reflection symmetry. CHR \cite{chr} took the natural choice of the 
five
dimensional bulk metric to be one describing a non-rotating black string
in the brane world. Inclusion of a three brane with reflection symmetry 
required the brane tension to be compatible with the junction conditions
relating the extrinsic curvatures on either side of the three brane. The
metric on the brane could then be recast into a standard Schwarzschild
metric with a
rescaled ADM mass and a curvature singularity at $r=0$ on the brane.
However there appeared a generic curvature singularity at $z=\infty$ the
AdS horizon.
Examination of the geodesics showed this to be a p-p curvature
singularity such that the curvature invariant 
diverges on bound state geodesics at finite $r$ but remains finite on
the non-bound ones which go to $r=\infty$.

In an earlier communication \cite {gs} we had generalized the construction
of CHR to study the occurence of a four dimensional Kerr metric describing
a rotating black hole on the three brane in a five dimensional RS brane
world. To this end we had considered, following CHR, a five dimensional
rotating black string in the RS brane world. This choice was compatible
with the junction conditions as in the non-rotating case. The metric on
the brane could be suitably rescaled to be recast in the form of a usual
four dimensional Kerr metric. This leads to a scaling of the
ADM mass and the angular momentum of the
resulting black hole in terms of the ratio of the AdS length scale to the
location of the three brane in the extra fifth dimension. The Kerr metric
on the brane posessed the usual features of inner and outer horizons and
an ergosphere. The square of the curvature tensor exhibits the usual ring
singularity. However as in the non-rotating case there was a generic
singularity at the AdS horizon. Study of the geodesics showed that as in
the non-rotating case the curvature squared diverged at the AdS horizon 
only along the bound state geodesics. It was suggested that the
singularity at the AdS horizon was also a p-p curvature singularity
however the exact determination of a parallely propagated orthonormal
frame for the non-diagonal metric seemed computationaly intractable
requiring the solution of three coupled PDE. Furthermore the rotating
black string was suggested to be subject to the Gregory-Laflamme
instability which would be relevant near the AdS horizon. However a
clear idea of a stable bulk metric requires further investigations
especialy in the light of the conclusions of \cite {greg,greg1,maeda}.
However explicit calculations in the light of \cite {greg1, maeda} are
absent for rotating axisymmetric metrics of which the Kerr metric
is a special case.

Following our construction for the four dimensional Kerr metric on the
brane it is natural to enquire if the higher dimensional Myers-Perry
generaliztion of the Kerr metric \cite {myers} may be obtained from
a higher dimensional brane world perspective. In particular it would
be interesting to describe the diverse horizon and singularity structures
of these metrics for multiple non-zero angular momentum
components which are possible in higher dimensions. As has been discussed
earlier for consistency the brane world scenario must be embedded in an
appropriate string theory. This necessitates higher dimensional
realizations of the RS scenario or some of its variants \cite {hd}. This
naturaly requires
a description of higher dimensional black hole and cosmological solutions
on the appropriate co-dimension brane in such higher dimensional brane
world models to reproduce lower dimensional General Relativity.

To this end we
consider a ( N+1) dimensional RS brane world with a single (N-1) brane 
(codimension one) and examine the possibility of obtaining
the Myers-Perry rotating N dimensional black hole on the world volume of
the (N-1) dimensional brane. Our earlier construction suggests that to
this end, it is necessary to consider a black string generalization of
the Myers-Perry metric in the (N+1) dimensional RS brane world. For
simplicity we first consider the case of a single non-zero spin component
which closely mimics the four dimensional solution. The metric for
the Myers-Perry rotating black string with a single non zero
angular momentum component is given from eqn. (1) as,
\bea
ds^2 &=& {l^2\over z^2}  \big[ -(1-{\mu\over {r^{N-5}\rho^2}})dt^2 + ( r^2
+
a^2
+{{\mu a^2}\over{r^{N-5}\rho^2}} sin^2 \theta )sin^2 \theta d\phi^2
\nonumber \\    
&+&{\rho^2\over\Delta} dr^2 + \rho^2d\theta^2 + r^2 cos^2\theta
d\Omega^{N-4}
 - {{2 \mu a}\over{ r^{N-5}\rho^2}}d\phi dt + dz^2\big ],
\eea.

The metric in eqn. (17) is
automaticaly a solution to the relevant Einstein equations for 
slice of the (N+1) dimensional AdS space, as the corresponding
N-dimensional Myers-Perry metric is Ricci flat\cite {myers}. This may be
recast into
the standard form as in eqn. (1) 
by suitable rescaling. The ADM mass and the angular momentum
for the rotating N-dimensional black hole on the brane are then given 
by rescaled values as
\be
M^*={l\over z_0} {{N-2 A_{N-2}}\over {16 \pi}} \mu \ \ \ 
J^{yx*}=\frac{l^2}{z_0^2}{A_{N-2}\over {8 \pi}}\mu a
=\frac{l^2}{z_0^2} {2\over {N-2}}M
a
\ee
It is seen from eqn. (17) that the horizon will occur for $g^{rr}=0$ and a
solution for this always exists for $N > 4$ for arbitrarily large angular
momentum. The square of the ( N + 1 ) dimensional curvature tensor is
given from eqn. (16). The N- dimensional Myers-Perry metric has been shown
in \cite {myers} to posess a curvature singularity at $r=0$ which occurs
at the edge of the (N-2) ball $x^{i^2} + y^{i^2} = a^{2}_i$. For a single
non-zero angular momentum this is the higher dimensional equivalent of the
ring singularity for the standard Kerr metric. The N+ 1 dimensional
curvature tensor naturaly inherits this singularity at $r=0$ on the (N-1)
brane.
However as
stated earlier a generic singularity appears at the AdS horizon as
$z\rightarrow
\infty $ which is obvious from eqn. (16). 

The metric for the Myers Perry black string being a stationary
axisymmetric metric posesses time like Killing isometries along the $t$
and $\phi^i$ directions where $i=1... {{N-2}\over 2}$. Considering only
a single
non-zero angular momentum component and restricting $\theta={\pi\over 2}$
we arrive at an effective four dimensional metric. In this case 
if $u$ is the tangent
vector to a timelike or null geodesic with an affine parameter $\lambda$,
the timelike Killing vectors $\xi={\partial \over{\partial t}}$ and
$\chi={\partial \over{\partial \phi}}$ gives rise to two conserved
quantities
$E=-\xi.u$ and $L=\chi.u$.
Rearrangement of these equations \cite{wald}
provides us with the geodesic equations for the $t$ and $\phi$ directions
for motion in the equatorial plane $\theta={\pi\over 2}$. These turn out
to be as follows
\begin{equation} \frac{dt}{d\lambda}= \frac{z^2}{l^2\Delta}\left[
\left(r^2+a^2+a^2 \alpha\right) E- a \alpha L\right]
\end{equation} \begin{equation}
\frac{d\phi}{d\lambda}=\frac{z^2}{l^2\Delta}\left[\left(1-\alpha\right)L
+a\alpha E\right] \end{equation} 
where $\alpha=\frac{\mu}{r^{N-3}}$.
The $z$ equation is then given as \cite {chr, gs}
\begin{equation}
\frac{d}{d\lambda}\left(\frac{1}{z^2}\frac{dz}{d\lambda}\right)=-
\frac{\sigma} {zl^2}. \end{equation} Here $\sigma=0,1$ for null and
the timelike geodesics respectively. The solution for the $z$ equation
is identical to the corresponding five dimensional case where we had
$z=$constant or 
\begin{equation} 
z=\frac{-z_{1}l}{\lambda},
\end{equation} 
for null geodesics and
\begin{equation}
z=-z_{1}cosec(\lambda/l). \end{equation}
for the timelike geodesics. The first solution is ignored as it describes
the Schwarzschild case.
The radial equation also remains identical
to the five dimensional case and is given by \cite {gs}
\begin{equation}
\left(\frac{dr}{d\lambda}\right)^2+\frac{z^4}{l^4}\left[\left(\frac{L^2-a^2E^2}
{r^2}-\frac{2M}{r^3}(aE-L)^2-E^2\right) \right ]+\frac{l^2}{z_{1}^2}
\frac{\Delta}{r^2} =0 \end{equation}
for both timelike and the null geodesics. This may be rescaled \cite
{chr,gs} to remove
any explicit $z$ dependence and after introduction of a new affine
parameter $\nu$,
leads to the standard radial equation for
timelike geodesic for a four dimensional Kerr metric which is given as
\cite
{wald}
\begin{equation}
\left(\frac{d\tilde{r}}{d\nu}\right)^2 + \left[\frac{\tilde{L}^2
-\tilde{a}^2\tilde{E}^2}{\tilde{r}^2}-\frac{2\tilde{M}}{\tilde{r}^3}
(\tilde{a}\tilde{E}-L)^2-\tilde{E}^2+\frac{\tilde{\Delta}}{\tilde{r}^2}\right]=0
\end{equation}
where $\nu$ is the proper time along the geodesic. Notice that
both the null and timelike geodesics for $\theta={\pi\over 2}$ in ( N+1)
dimensions reduces to four dimensional
time like geodesics with a consequent relationship between the two affine
parameters.
The conclusions
obtained in \cite {gs} for the behaviour of the solution near the
singularity are unchanged in this case. In particular the curvature
squared near
the AdS horizon
diverges
along the four dimmensional bound geodesics but remains finite along
non-bound 
ones. So we see that the Myers-Perry black string with a single
\nzam\  in the (N+1) dimensional brane world leads to a behaviour
similar to the five dimensional case \cite {gs} for $\theta=\frac{\pi}{2}
$. This is expected,  as in this limit we have an effectively four
dimensional metric in the equatorial hyperplane. In the next section
we describe a general N dimensional \myp
solution with multiple
\nzam\  on the (N-1) brane from the (N +1) brane world perspective. 

\section{Myers-Perry Black String with Multiple Spins in a Brane World}

Having obtained a description of the N dimensional \myp \ solution with
a single \nzam \ on the N-1 brane, we now turn to the most general
solution
with multiple non-zero angular momentum components. To this end we
consider a black string
generaliztion
of the rotating \myp solution with multiple \nzam \ given by eqn. (3) in a
(N+1) RS
brane world. The metric for this black string in the Boyer-Lindquist
coordinates is given as in eqn. (3),
\bea
ds^2 &=& \frac {l^2}{z^2}\big [ -dt^2 +\sum_k (r^2 + a_k^2)(d\mu_k^2 +
\mu_k^2d\phi_k^2)\nonumber \\
&+& {{mr^{2-\epsilon}}\over \Pi F}\big ( dt - \sum_k a_k\mu_k^2
d\phi_k\big )^2 \nonumber \\
&+& {{\Pi F}\over {\Pi -mr^{2-\epsilon}}} dr^2
+\epsilon r^2 d\alpha^2 + dz^2\big ]
\eea
in the coordinates and notations of the earlier section. This may be
recast
into the standard form cf. eqn (3) by suitable rescalings as earlier.
The ADM mass and the angular momentum are given by the rescaled
values as before;
\be 
M^*={l\over z_0} {{N-2 A_{N-2}}\over {16 \pi}} \mu\ \ \ 
J^{y^{i}x^{i}*}=\frac{l^2}{z_0^2}{A_{N-2}\over {8 \pi}}\mu a_i
=\frac{l^2}{z_0^2} {2\over {N-2}}M a_i
\ee
where $z=z_0$ is the location of the (N-1) brane.

The constant $r$ surfaces
for odd ( N-1) spatial coordinates are now products of $\frac {S^1}{Z_2}$
with (N-2) ellipsoids. The intersection of these with $x^i=y^i=0\
, i>1$ gives the familiar two dimensional surfaces of the standard Ker
metric but are now extended along the ( N +1) th direction. From \cite
{myers}
these
surfaces are now product of $\frac {S^1}{Z_2}$ with $S^2$ squashed along
$w$
direction or product of ellipsoids about $w$ with $\frac {S^1}{Z_2}$. As
earlier the extended ellipsoids now intersect $w$-axis at $w=r$ and the
$(x^1,y^1)$ plane at the circle $x^{1^2} + y^{1^2}=r^2 + a^{1^2}$ and the
$r=0$ surface degenerates to the product of $\frac {S^1}{Z_2}$ with the
disk of radius $a_1$ on the $(x^1,y^1)$ plane. The entire surface in this
case is described as a product of  $\frac {S^1}{Z_2}$ with a squashed
(N-2) sphere with rotational symmetry in
each of the $(x^i,y^i)$ plane and the $r=0$ surface degenerates to a
product of  $\frac {S^1}{Z_2}$ with an (N-2p-2) ball where $p$ is the
number of vanishing angular momentum components $a_i$. Exactly simmilar
consideration holds for even ( N-1) spatial coordinates where the 
$r=0$ surface degenerates to the product of $\frac {S^1}{Z_2}$ with an
(N- 2p -1) ball if $p\neq0$. 
The horizons for this \myp\ black string metric are now extended along
the ( N+1) th direction with the topology $S^{N-2}\times R\times \frac
{S^1}{Z_2}$. They are given by the same conditions as before
cf. eqn. (12) with rescaled parameters as in eqn. (26).  

As stated earlier in section two, the \myp\  metric in N dimensions
posesses a 
curvature singularity at $r=0$. Evaluation of the curvature tensor in
an orthonormal frame which feels the local tidal forces shows this
divergence apart from a few exceptional cases where the $r=0$ does not
appear to be entirely singular and needs extension to $r < 0$. For a
specific component $R_{uvuv}\sim r^{-2p-1}$ for odd (N-1)
spatial coordinates and  $R_{uvuv}\sim r^{-2p}$ for even (N-1) with
$p\geq 1$. From eqn. (16) which gives the (N +1) dimensional curvature
squared it is obvious that the full ( N+1) dimensional black string metric
in the corresponding brane world has a singularity at $r=0$. However eqn.
(16)
also shows that 
\be
R_{pqrs}R^{pqrs}\sim \frac{z^4}{r^{2(2p-\epsilon)}}
\ee 
apart from a N dependent factor,
which exhibits a singularity at $z=\infty$, the AdS horizon. As mentioned
earlier such a singularity seems to originate from the linearized
approximation.

To further examine the nature of this singularity at the AdS horizon
we investigate the geodesic equations. For this purpose we specialise
to the case of odd (N-1) spatial coordinates for convenience. Similar
method will hold also for the case of (N-1) even.
The black string metric in the
(N+1) dimensional AdS space posesses Killing isometries corresponding to
time translations and rotations in the paired coordinate planes $x^{2k-1}-
x^k$ with $\phi^k$ being the angles in these planes. The Killing vectors
$(\xi , \chi_k )$ are thus  $\partial_t$ and $\partial_{\phi_k}$.
Considering $u$ to be the
tangent vector to a timelike or null geodesic parametrized by the affine
parameter $\lambda$ the timelike Killing vectors gives rise to the
following conserved quantities
\be
E=-\xi.u \ \ \ \ L_i=\chi_i.u
\ee
From eqn. (3) which describes the black string metric, we restrict
the polar angles $\theta_i=\frac {\pi}{2}$ and arrive at a reduced metric
of the $\frac{N + 2}{2}$ dimensional equatorial hyperplane of the (N +1)
dimensional space-time as
\bea
ds^2&=& \frac {l^2}{z^2}\big [ -dt^2 +\sum_k (r^2 + a_k^2)
d\phi_k^2 \nonumber \\
&+& {{mr}\over \Pi F}\big ( dt - \sum_k a_k
d\phi_k\big )^2 
+{{\Pi F}\over {\Pi -mr}} dr^2
+ dz^2\big ]
\eea  where $\epsilon=1$ for the case under discussion.
Using this metric the conserved quantities corresponding to the timelike
Killing isometries may be established as 
\def\epsa {2-\epsilon}
\be
E=\frac{l^2}{z^2}\big [ (1-\frac {\mu r}{\Pi F})\frac
{dt}{d\lambda}
 +\frac {\mu r}{\Pi F}\sum_k a_k\frac {d \phi_k}{d\lambda}\big ]
\ee
and
\be
L_i=\frac{l^2}{z^2}\big [ -\frac {\mu r}{\Pi F} a_i\frac
{dt}{d\lambda}
    + (r^2 + a_i^2 + \frac {\mu r}{2 \Pi F} a_i^2)\frac
{d\phi_i}{d\lambda} + \frac {\mu r}{2 \Pi F} \sum_{j\neq
i}a_i a_j \frac{d\phi_j}{d \lambda} \big
]
\ee
This is a system of $\frac {N}{2}$ linear equations in $\dot {t}$ and
$\dot {\phi_k}$ where the dot denotes derivative with respect to the
affine
parameter. Rewriting this system as a matrix equation we have with
$\dot{t},\dot{\phi_k} = q^{\alpha}$ and $E, L_i = c_{\beta}$ where $\alpha
,\beta = 1...\frac {N}{2}$,
\be
\frac{l^2}{z^2}\sum_{\beta}M_{\alpha \beta}q_{\beta}=c_{\alpha}
\ee
where $M_{\alpha \beta}$ is the matrix of the coefficients and the
conformal factor of the metric has been separated out. Assuming this
to be non-singular so that a solution exists, we have
\be
q_{\alpha}= {z^2\over l^2}\sum_{\delta}(M)^{-1}_{\alpha \delta}c_{\delta}
\ee
Here $M_{\alpha \beta}$ is a function of $r$ and $a_k, \mu$ are
parameters.
The geodesic equation for the $z$ coordinate which denotes the AdS
direction remains the same as in eqn. (21) with the same solution
for the timelike and the null geodesics. Using these solutions and
eqn.(33)
we may obtain the radial geodesic equation for the rotating black string
metric in the $\frac{N+2}{2}$ equatorial hyperplane by substitution in 
\be
-\sigma=g_{mn}u^m u^n
\ee
where $\sigma =0, 1$ for null and timelike geodesics respectively.
The radial equation for a timelike geodesic thus obtained is given as
\bea
\frac{\Pi F}{\Pi -\mu r}\dot{r}^2 + \frac{z^4}{l^4}\big [
\sum_{\alpha}\sum_{\beta} (M_{\alpha \beta})^{-1}c_{\alpha}c_{\beta} +
\frac{l^2}{z_1^2}\big ]=0 
\eea
To specialise we choose a value of N in this case for which a
consistent
horizon and singularity structure exists for the Myers-Perry metric.
For the case of (N-1) odd spatial dimensions the first such non-trivial
value is
$N=6$ 
. Hence we consider a (6+ 1)
dimensional RS brane world with a 5 dimensional brane for which we have
$E, L_1, L_2$ as the
three constants of motion.
For this scenario it is easy to obtain the explicit form of the matrix
$M_{\alpha\beta}$ and solve for the constants thus arriving at
the radial geodesic equation
in the equatorial hyperplane for this case 
Following CHR and our
earlier work \cite {gs} for the Kerr black hole we may introduce a new
affine
parameter $\nu$ and simultaneously rescale all the coordinates, mass,
parameters and the constants of motion to remove the explicit $z$
dependence from the radial equation and obtain an effective 
radial geodesic equation in N=6 dimensions. This equation is given as
\bea
\big ({d{\tilde r}\over {d\nu}} \big ) + V({\tilde r}) =0
\eea
where $V({\tilde r})$, the effective potential is an involved function of
${\tilde r}$ whose
explicit form is not very
instructive. Except the fact that $V({\tilde r})$ contains ${\tilde r}^2$
as the maximum
power of ${\tilde r}$ in the denominator ruling out any stable bound
orbits. So the
only allowed geodesic orbits are non-bound ones.
The
asymptotic
behaviour of the corresponding time-like radial non bound geodesics would
then correspond to late time behaviour or $\nu\rightarrow \infty$ as
$\nu$ would be the proper time. Thus
these geodesics reach the AdS horizon at $z=\infty$. Hence the
curvature squared from eqn. (28) remains finite along such geodesics
signaling the
presence of a p-p curvature singularity at the AdS horizon. To explicitly
illustrate this it is necessary to obtain the curvature components in
an orthonormal frame parallely propagated to $z=\infty$. However even in
the Kerr case the explicit determination becomes computationaly
intractable \cite {gs} and more so in these higher dimensions. Hence we
will not attempt an explicit determination in this case either but
emphasize that such an ON frame should clearly exist.

The question of the stability of the rotating black string solution is 
still an unsolved problem and an explicit calculation in the light of
\cite {greg1, maeda} has to be done for the axisymmetric solutions
considered here. Two possibilities abound namely that the metric
pinches off due to instabilities before reaching the AdS horizon or
collapses to an
intermediate stable solution which is non-singular everywhere except
on the brane. However it has been shown that in (2 +1) dimensions
where exact AdS C metrics are available the brane world solution is
nonsingular at the AdS horizon. This suggests that the pathological
singularity at the AdS horizon is simply an artifact of the linearized
approximation in the RS brane world scenario.

\section {Summary and Conclusions.} 

To summarize we have 
extended our earlier work describing
a four dimensional Kerr solution on a three brane in a five dimensional RS
brane world to a higher dimensional brane world. The motivation for such
an exercise, as mentioned earlier follows from the fact that for
consistency the brane world models must be embedded in an appropriate
string theory. This requires generlaizations of these models to higher
dimensions. Furthermore to confirm that the usual predictions of lower
dimensional General Relativity on the brane world volume are consistent
requires investigation of known black hole and cosmological solutions
in higher dimensions. The absence of exact C metrics in $D>4$ requires
a linearized approximation for such studies and the method of CHR seems
to be a suitable one in such a framework. 

In particular we have
considered a
(N+1) dimensional RS brane world with a (N-1) brane and a 
Myers-Perry black string with an extended singularity. Such a
choice is necessitated
for compatibility with the junction conditions at the location of the
(N-1) brane.
The N dimensional metric on the (N-1) brane in
this case describes a rotating black hole with a Myers-Perry metric.
The cases of single and multiple \nzam\ has been dealt with separately.
The single \nzam\ case closely mimics the corresponding four dimensional
Kerr solution on a three brane starting with a rotating black string
in a five dimensional brane world. The singularity on the brane
and the geodesic equations for the equatorial hyoerplane too are
identical while the singular asymptotic behaviour at the AdS horizon are
simmilar to the four dimensional case. As shown in \cite {myers} the
metric with multiple \nzam\ 
exhibits a diverse horizon and singularity structure which also depends
on whether the number of spatial corrdinates are even or odd. Starting
with such a multiple \nzam\ black string we have been able to describe
this horizon and singularity structure from a brane world
perspective. Furthermore we
have obtained the geodesic equations for the equatorial hyperplane in a
closed form for arbitrary N
and multiple \nzam. As an example we have chosen a specific value $N=6$
which is the first non-trivial dimension for which the Myers-Perry metric
has a consistent horizon and singularity structure and obtained an
explicit form of the radial geodesic equation. It is obvious from the
effective potential that stable bound geodesic orbits are ruled out in
this higher dimensional case. Analysis of the radial equation and the
non-bound geodesic orbits clearly indicates the existence of a p-p
curvature singularity at the AdS horizon.

It is important to consider the question of the stability of our solution
in the light of \cite {greg1, maeda} where the corresponding analysis have
been performed for the spherical symmetric Schwarzschild case. Such a
generaliztion will possibly lead to a more complete understanding of the
asymptotic behaviour for these solutions. Another related issue is the
determination of the exact off-brane bulk metric for four or more
dimensions. The description of the Kerr-Newman black hole on the brane
for the five dimensional model is also an outstanding issue. This is
not relevant to our solution as no charged generalization of the
Myers-Perry solution has been established. Some of these issues are
being currently studied. It is also important to investigate the
possibility of a generaliztion of the AdS C metrics to higher dimensions
so that exact calculations as in dimensions less than 5 may be performed.
Hopefuly some of these issues will be clarified in the near future.

{\bf Acknowledgements :} 

I would like to thank Sudipta Mukherji for bringing reference \cite
{myers} to my attention, Naresh Dadhich for comments and Yogesh Kumar
Srivastava for collaboration
in the initial part. Further I would like to acknowledge Koushik Ray,
Sreerup Raychaudhuri and other members of
of the High Energy Physics Journal Club at IIT Kanpur for useful
discussions and Santosh Kumar Rai for computational help.
\vfil
\eject
%%%%%%%%%%%%%%%%%%%%%%%%%%%%%%%%%%%%%%%%%%%%%%%%%%%%%%%%%%%%%%%%%%
%\Newpage

\end{document}